\documentclass[aps,pra,twocolumn,amsmath,amssymb,floatfix,showkeys,showpacs,reprint]{revtex4-1}
\usepackage{graphicx}
\usepackage{bm} % bold math
\usepackage{color}
\usepackage{graphics} %  <------- for LatTeX + PS
\usepackage{graphicx} % Include figure files
\usepackage{epsfig}
\usepackage{amssymb}
\usepackage{amsmath} %,amssymb}
\usepackage{amsthm}
\usepackage{amstext}
\usepackage{dcolumn} % Align table columns on decimal point
\usepackage{makeidx}
\usepackage{subfigure}
\usepackage{float}

\begin{document}

\title{The possibility of direct observation of the Bloch--Siegert shift in coherent dynamics of multiphoton Raman transitions}

\author{A.P. Saiko$^{1}$}\email{saiko@physics.by}
\author{S.A. Markevich$^{1}$}
\author{R. Fedaruk$^{2}$}
\affiliation{$^{1}$Scientific-Practical Material Research Centre, Belarus National Academy of Sciences, 19 P.Brovka str., Minsk 220072 Belarus}
\affiliation{$^{2}$Institute of Physics, University of Szczecin, 15 Wielkopolska str., 70-451, Szczecin, Poland}

\date{\today}
\begin{abstract}
We study Rabi oscillations of the second-order Raman transition realized on dressed states of a qubit excited by an amplitude-modulated microwave field. The co-rotating component of the ultrastrong low-frequency modulation field excites virtual multiple photon processes between the dressed states and forms the Rabi frequency in the so-called rotating wave approximation (RWA). The counter-rotating modulation component also gives a significant contribution to the Rabi frequency owing to the Bloch--Siegert effect. It is shown that for properly chosen parameters of the modulation field and qubit, the Rabi oscillations in the RWA vanish due to destructive interference of multiple photon processes. In this case the Rabi oscillation results exclusively from the Bloch--Siegert effect and is directly observed in the time-resolved coherent dynamics as the Bloch--Siegert oscillation. Correspondingly, in Fourier spectra of the coherent response, triplets are transformed into doublets with the splitting between the lines equal to twice the Bloch--Siegert shift. We demonstrate these features by calculations of the qubit's evolution in the conditions of  experiments with a NV center in diamond, where Raman transitions were observed. The direct observation of the Bloch--Siegert oscillation offers new possibilities for studying driven quantum systems in the ultastrong regime.
\end{abstract}
\pacs{42.50.Ct, 42.50.Dv, 42.50.Hz, 76.30.Mi}

\maketitle
\section{INTRODUCTION}

In two-level quantum systems driven by a strong oscillating electromagnetic field, the Bloch--Siegert effect \cite{p1} is a phenomenon in quantum physics in which the observed resonance frequency is changed by the presence of a counter-rotating (off-resonance) component of the driving field. This shift of the resonance frequency is usually negligible for optical transitions \cite{p2}, but becomes significant for precision nuclear magnetic resonance (NMR) experiments \cite{p3}, preventing accurate observation of the resonance frequencies. In the dispersive regime this shift sometimes is referred to as dynamical Stark shift \cite{p4}. The appearance of the Bloch--Siegert shift means that the commonly used rotating wave approximation (RWA) is broken and the contribution of the counter-rotating (non-RWA) terms to the coupling Hamiltonian must be taken into account.

In the past decade, studies of the resonant matter-light interaction evolve toward the ultrastrong ($0.1<g/\varepsilon\lesssim1$) and deep strong ($g>\varepsilon$) coupling regime where the coupling strength $g$ is comparable to, or exceeds, the transition frequency $\varepsilon$ between two energy levels of the quantum system. As a result, the contribution of the non-RWA terms results in complex dynamics of the field-matter interaction (see, e.g. \cite{p5, p6, p7*, p8*}) and makes difficulties for its analytical description.

The coherent dynamics of resonant interaction between coherent radiation and two-level systems (qubits) can be described in terms of Rabi oscillations between the two energy eigenstates. This dynamics is extremely important for quantum information processing \cite{p7}, protection against decoherence \cite{p8} and is widely studied for various quantum objects such as nuclear and electronic spins \cite{p3, p7}, natural atoms \cite{p2}, artificial atoms such as quantum dots \cite{p9} and superconducting qubits \cite{p10}. The rate of coherent manipulations of qubit's states is characterized by the Rabi frequency and depends on the strength of the driving field. The increase in the manipulation rate results in faster state operation, however, can lead to the strong driving regime, where, due to breakdown of the RWA, complex dynamics of qubits occurs. The steady-state response of quantum systems, mainly superconducting qubits, under their strong continuous-wave driving have been studied \cite{p10, p11}. The strong driving of qubits has been investigated in Landau-Zener-St\"{u}ckelberg interferometry on quantum dots \cite{p12, p13} and in hybrid quantum systems (a superconducting flux qubit -- a single nitrogen-vacancy (NV) center \cite{p14}). Recently, in the ultrastrong regime ($g>\varepsilon$), the time-resolved Rabi oscillations of artificial atoms including superconducting flux \cite{p15, p16, p17} and charge \cite{p18} qubits as well as a single NV center in diamond \cite{p19, p20}, dressed states of NV centers \cite{p21}, nuclear spins \cite{p22} and a single NV center under mechanical driving \cite{p23} have been studied. The presence of the various frequency components in the Rabi oscillations and the Bessel-function dependence of the quasienergy difference on the driving strength have been observed \cite{p16, p17, p18}. In particular, the ultrastrong regime has been realized in the solid state system for the quantum Rabi model \cite{p25, p26, p27*, p28*}, where the light field is quantized, and the semiclassical Rabi model \cite{p17, p19, p20}, where light is treated as an external classical control field. The deep strong regime of the quantum Rabi model has been considered in \cite{p24}. Note that the extremely strong driving (for the semiclassical Rabi model) induces the transitions between the qubit's levels not only at the resonant or near resonant excitation, but also when the frequency of driving field is far away from resonance and exceeds significantly the qubit transition frequency \cite{p27, p28}.

In terms of the dressed\textit{-}state formalism \cite{p29}, the dressing of qubit by the resonant electromagnetic field gives rise to new energy levels of the coupled field-qubit system and the Rabi frequency characterizes the splitting of each bare level. The applied second field with the frequency closed to the Rabi frequency excites effectively transitions between the dressed states. This phenomenon, called the Rabi resonance, has been observed for spin ensembles \cite{p30, p31} and a single spin \cite{p32} in electron paramagnetic resonance, NMR \cite{p33, p34, p35, p36} as well as for atoms in the optical range \cite{p37, p38, p39}. Additional resonances occur at the subharmonics of the Rabi frequency \cite{p35, p36, p37, p40}. The additional Rabi oscillations between the dressed states at the frequency determined by the strength of the second driving field have been observed under the Rabi resonance condition \cite{p30, p31, p32, p34, p36, p38, p39}. Since the Rabi frequency in the second driving field can easily be obtained even larger than the Rabi frequency in the first driving field, the RWA is often broken in a description of the coherent dynamics of the dressed-state transitions and under such strong driving the Bloch--Siegert effect becomes significant \cite{p31, p32, p41}. In the time-resolved observations, the Rabi resonance was realized when the first driving field was in resonance with the two-level system. In this case the bichromatic control of Rabi oscillations between doubly dressed states and prolongation of their coherence can find applications in quantum information processing \cite{p42, p43, p44}. Moreover, in a single-spin magnetometry, Rabi oscillations between doubly dressed states open the possibility for the direct and sensitive detection of weak radio-frequency magnetic fields \cite{p45, p46}.

Recently, rich Floquet dynamics has been demonstrated \cite{p47}. So-called Floquet Raman transitions have been observed in the solid-state spin system of NV center in diamond driven by the microwave field with its low-frequency amplitude modulation \cite{p47}. The microwave frequency was detuned from the resonant frequency of the two-level system and Raman transitions between dressed spin states were excited by the low-frequency field when multiphoton Rabi resonances (termed also Floquet resonances \cite{p48}) were realized. The observe dynamics offers new capabilities to achieve effective Floquet coherent control of a quantum system with potential applications in quantum technologies or as a quantum simulator for the physics of periodically driven systems \cite{p47}. Closed-form expressions for the Rabi frequencies of Raman transitions have been obtained beyond the RWA for the low-frequency driving component \cite{p49}. It was shown that the ultrastrong regime ($g/\varepsilon \approx 0.2$) for the semiclassical Rabi model is reached in the experiment \cite{p47} resulting in the significant Bloch-Siegert shift. However, it still remains an interesting possibility to achieve the stronger light-matter coupling and observe so far unexplored behaviors of Rabi oscillations for Raman transitions. In particular, in the experiment \cite{p47} as well as in experiments, where the counter-rotating terms manifest important impacts, the Bloch-Siegert effect does not observe separately from other oscillating processes in the qubit's dynamics. To our knowledge, there are no propositions for filtering the Bloch-Siegert shift in literature and its direct observation remains a challenge.

In the present paper, we focus on the ultrastrong regime of the coherent dynamics of the second-order Raman transition excited by the amplitude-modulated microwave field in the two-level system and propose the method for direct observation of the Bloch-Siegert oscillation. By using the semi-classical Rabi model and in the framework of the nonsecular perturbation theory based on the Bogoliubov averaging method, we obtain a closed-form expression for the Rabi frequency of this transition beyond the RWA.  It is demonstrated that at the ultrastrong modulation field, due to destructive interference of multiple photon processes, the Rabi oscillations disappear under some ratio of the amplitude and frequency of the modulation field. At this condition the oscillation resulting from the Bloch--Siegert effect is only observed. Note that usually the Bloch--Siegert effect does not observe directly, but it is revealed as the shift of the resonant frequency \cite{p3} or the complex (multifrequency) character of observed Rabi oscillations \cite{p22}. The analytical description of coherent dynamics of the second-order Raman transition is presented in Sec. II. Time and spectral manifestations of the Bloch--Siegert effect are demonstrated in Sec. III using the calculations at the parameters of the driving field which can be realized in experimental studies of NV center in diamond \cite{p47}. The effects of the phase of the low-frequency modulation field are also considered.

\section{COHERENT DYNAMICS OF THE SECOND-ORDER RAMAN TRANSITION}
Raman transitions between Floquet dressed states of an initially two-level spin system are excited by a microwave field $V(t) = \Delta _{x} \cos (\omega _{d} t)+2A\cos (\omega _{d} t)\sin (\omega t+\psi)$, where $\cos (\omega _{d} t)$ describes the high-frequency component of the field, $\sin (\omega t+\psi)$ represents the low-frequency component with the initial phase $\psi$, and the amplitudes of these componentes $ \Delta _{x}$, $A\ll\omega _{d}$ \cite{p47}. The Hamiltonian of the two-level system at such driving can be written as $H_{lab} = \frac{\Delta E}{2} \sigma ^{z} +\Delta _{x} \cos (\omega _{d} t)\sigma ^{x} +2A\cos (\omega _{d} t)\sin (\omega t+\psi)\sigma ^{x}$, where $\Delta E$ is the transition energy between the ground and excited levels; $\sigma ^{z}$ and $\sigma ^{z}$ are Pauli operators. In the frame rotating with the driving field frequency $\omega_d$ and in the RWA for this field (since $\omega, \Delta _{x}, A \ll \omega _{d}$), the Hamiltonian is
\begin{equation} \label{eq_01}
H=\frac{\Delta _{z} }{2} \sigma ^{z} +\frac{\Delta _{x} }{2} \sigma ^{x} +A\sin (\omega t+\psi )\sigma ^{x},
\end{equation}
where $\Delta _{z} =\Delta E-\omega _{d} $. The dynamics of the system  is described by the Liouville equation for the density matrix $\rho $: $i\partial \rho /\partial t=H\rho $ (in the following we take $\hbar =1$). Rotating the frame around the \textit{y} axis by angle of $\theta $ ($\rho \to \rho _{1} =U_{1}^{+} \rho U_{1} $, $U_{1} =e^{-i\theta \sigma ^{y} /2} $, and $\sigma ^{y} =(\sigma ^{+} -\sigma ^{-} )/i$), we can write the same equation with the Hamiltonian $H_{1} =U_{1}^{+} HU_{1} =\frac{\omega _{0} }{2} \sigma ^{z} +A\cos \theta \sin (\omega t+\psi )\sigma ^{x} +A\sin \theta \sin (\omega t+\psi )\sigma ^{z} $, where $\omega _{0} =\sqrt{\Delta _{z}^{2} +\Delta _{x}^{2} } $, $\sin \theta =\Delta _{x} /\omega _{0} $, $\cos \theta =\Delta _{z} /\omega _{0} $. After the second canonical transformation $\rho _{1} \to \rho _{2} =U_{2}^{+} \rho _{1} U_{2} $ with $U_{2} =\exp \left\{-i\left[\omega _{0} t-\frac{2A\sin \theta }{\omega } \cos (\omega t+\psi )\right]\frac{\sigma ^{z} }{2} \right\}$, we obtain the Liouville equation for $\rho _{2} $ with the Hamiltonian
$$H_{2} =U_{2}^{+} H_{1} U_{2} -iU_{2}^{+} \frac{\partial U_{2} }{\partial t} =$$
$$=\frac{A}{2i} \cos \theta \left[\sigma ^{+} \sum _{n=-\infty }^{\infty }J_{n} (a)e^{-in\pi /2}\right.\times $$
\begin{equation} \label{eq_02}
\times\Biggl. \left(e^{i(n+1)\omega t} e^{i(n+1)\psi } -e^{i(n-1)\omega t} e^{i(n-1)\psi } \right)e^{i\omega _{0} t} +H. c. \Biggr],
\end{equation}
where $J_{n} (a)$ is the Bessel function of the first kind and $a=2A\sin \theta /\omega $.
\begin{figure*}[ht]
\centering
\includegraphics[]{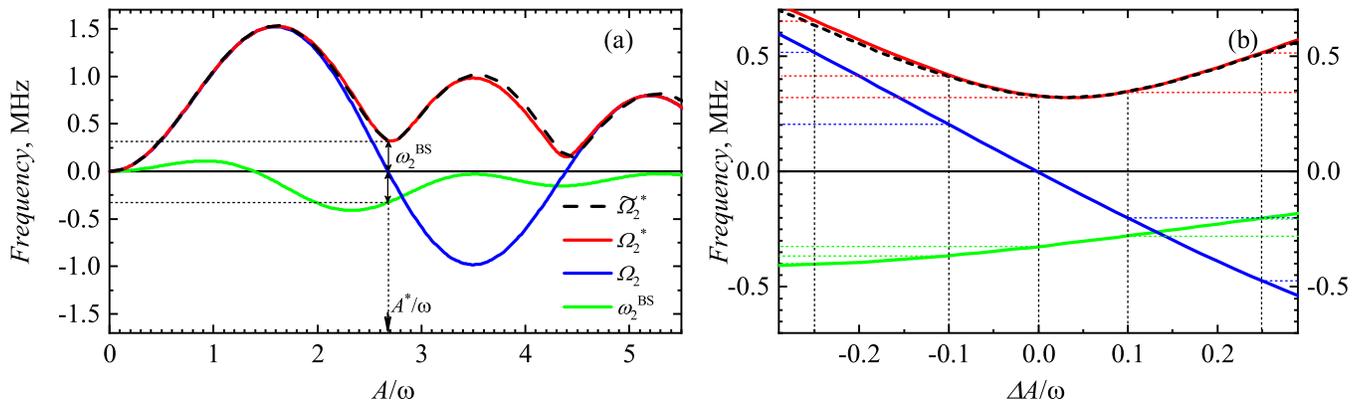}
\caption{(a) The dependence of the RWA Rabi frequency $\Omega _{2}$, the non-RWA Rabi frequency $\Omega _{2}^{*}$, $\tilde{\Omega }_{2}^{*}$ and the Bloch--Siegert shift $\omega _{2}^{BS}$ on the normalized amplitude of the low-frequency driving field at $\omega /2\pi$ = 5.22 MHz, $\Delta x/2\pi$ = 10 MHz, and $\Delta z/2\pi$ = 3 MHz. (b) The frequencies presented in (a) near $A^{*} /\omega$ show in more detail. This plot is useful to obtain the values of these frequencies for $\Delta A/\omega$ used in the following figures.}
%%\label{fig1}
\end{figure*}
\begin{figure*}[ht]
\centering
\center{\includegraphics[]{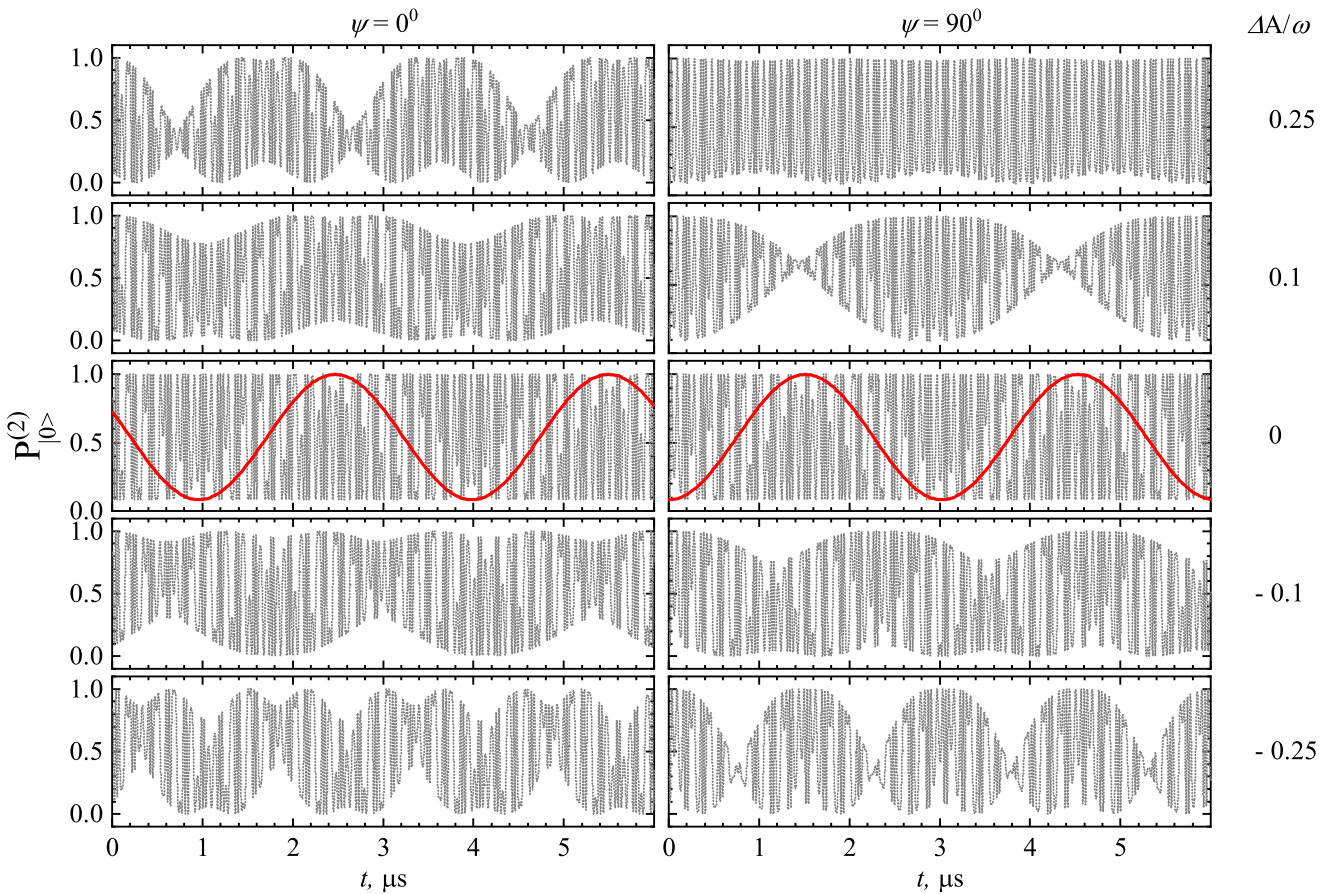}}
\caption{The state population of the spin level $|0 \rangle$ for the second-order Raman transitions as a function of the evolution time at $\omega /2\pi$ = 5.22 MHz, $\Delta x/2\pi$ = 10 MHz, and $ \Delta z/2\pi$ = 3 MHz. The strength of the low-frequency driving field is $A = A^{*} +\Delta A$, where $A^{*} /\omega$ = 2.68 and $\Delta A/\omega$ = 0.25, 0.1, 0, --0.1, --0.25. The coherent oscillations in the qubit's evolution are presented for the phase of the driving field $\psi$ = 0 (left panel) and $90^{0}$ (right panel). The red line shows the Bloch--Siegert oscillation.}
%%\label{fig2}
\end{figure*}
\begin{figure*}[ht]
\begin{minipage}[h]{0.47\linewidth}
\nonumber\center{\includegraphics[width=1\linewidth]{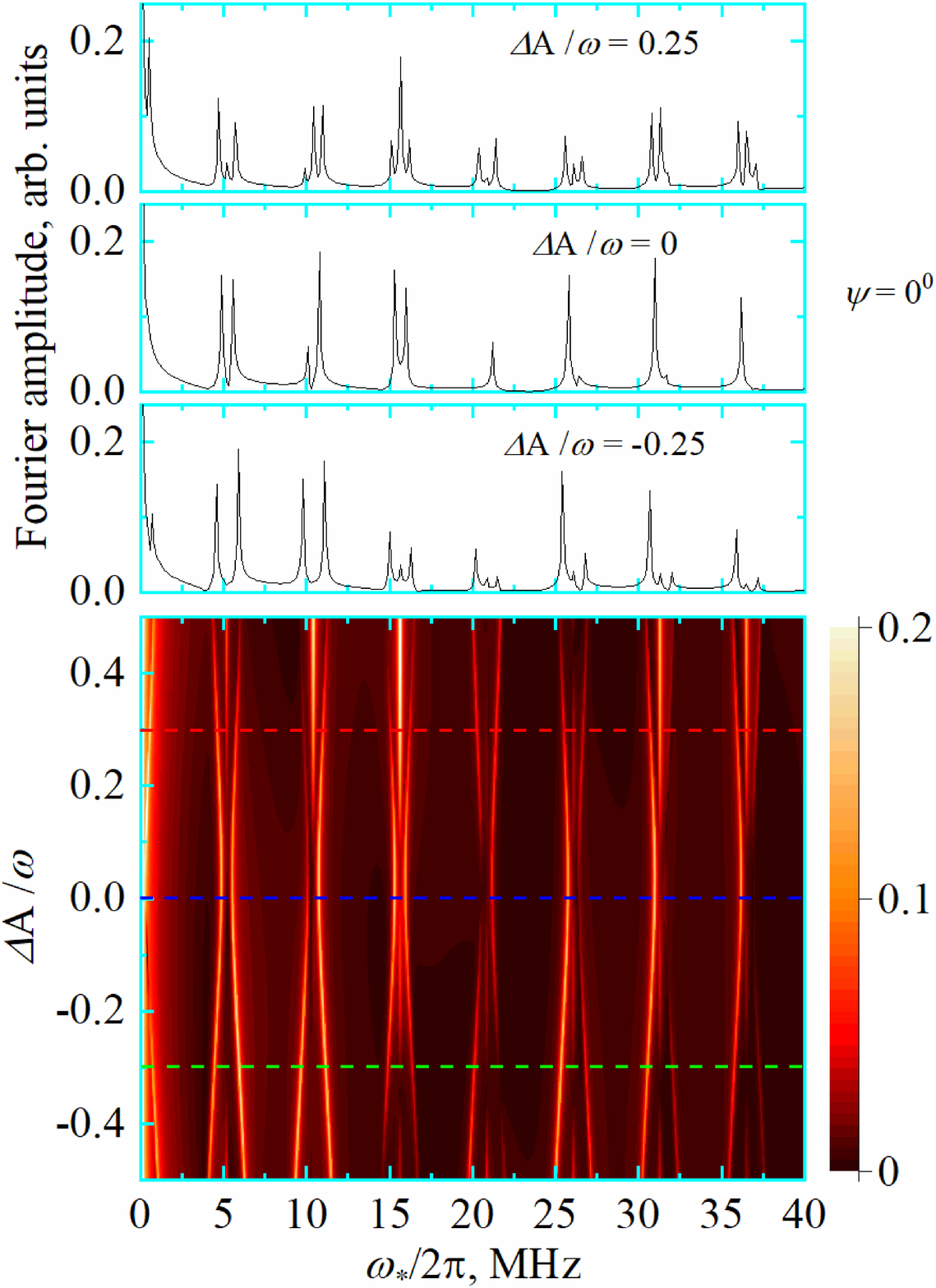}} \\
\end{minipage}
\hfill
\begin{minipage}[h]{0.47\linewidth}
\center{\includegraphics[width=1\linewidth]{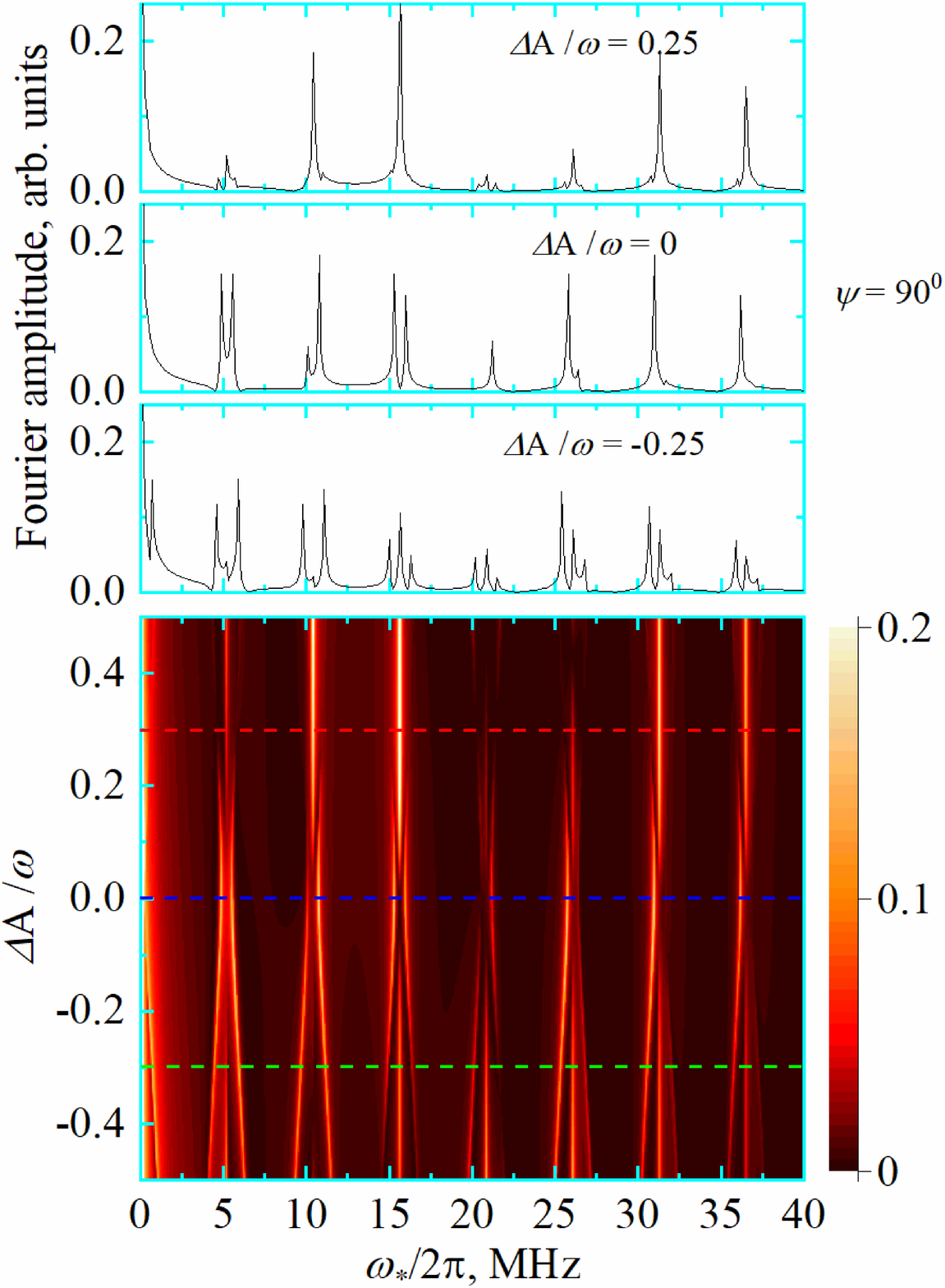}} \\
\end{minipage}
\caption{Fourier spectra as a function of $\Delta A/\omega = (A-A^{*})/\omega$ for the phase of the low-frequency driving field $\psi$ = 0 (left panel) and $90^{0}$ (right panel). Red, blue and green lines show cuts at $\Delta A/\omega$ equal to 0.25, 0 and --0.25, respectively. The other parameters are the same as in Fig. 2.
}
%%\label{fig3}
\end{figure*}

\begin{figure*}[ht]
\begin{minipage}[h]{0.47\linewidth}
\center{\includegraphics[width=1\linewidth]{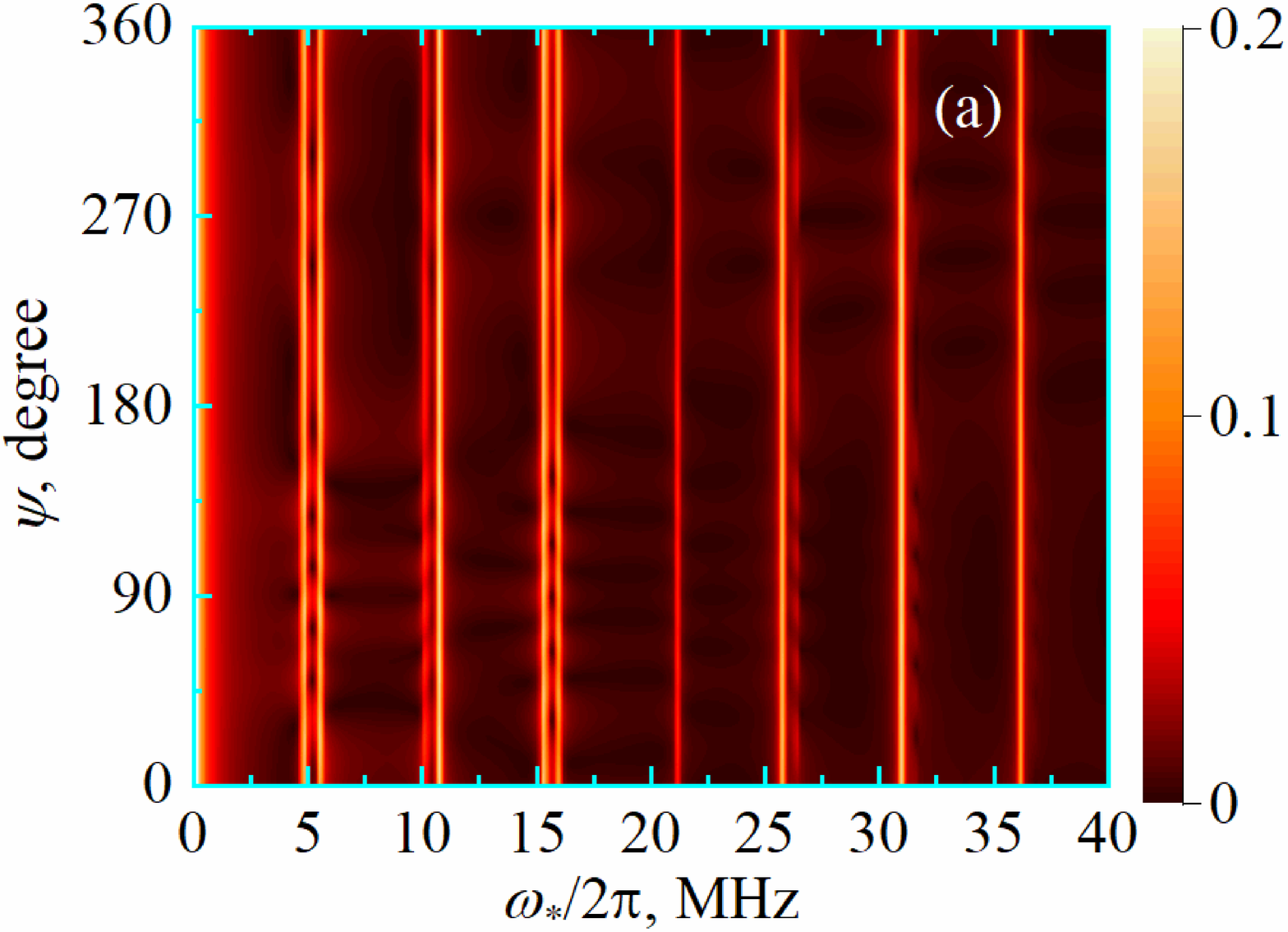}} \\
\end{minipage}
\hfill
\begin{minipage}[h]{0.47\linewidth}
\center{\includegraphics[width=1\linewidth]{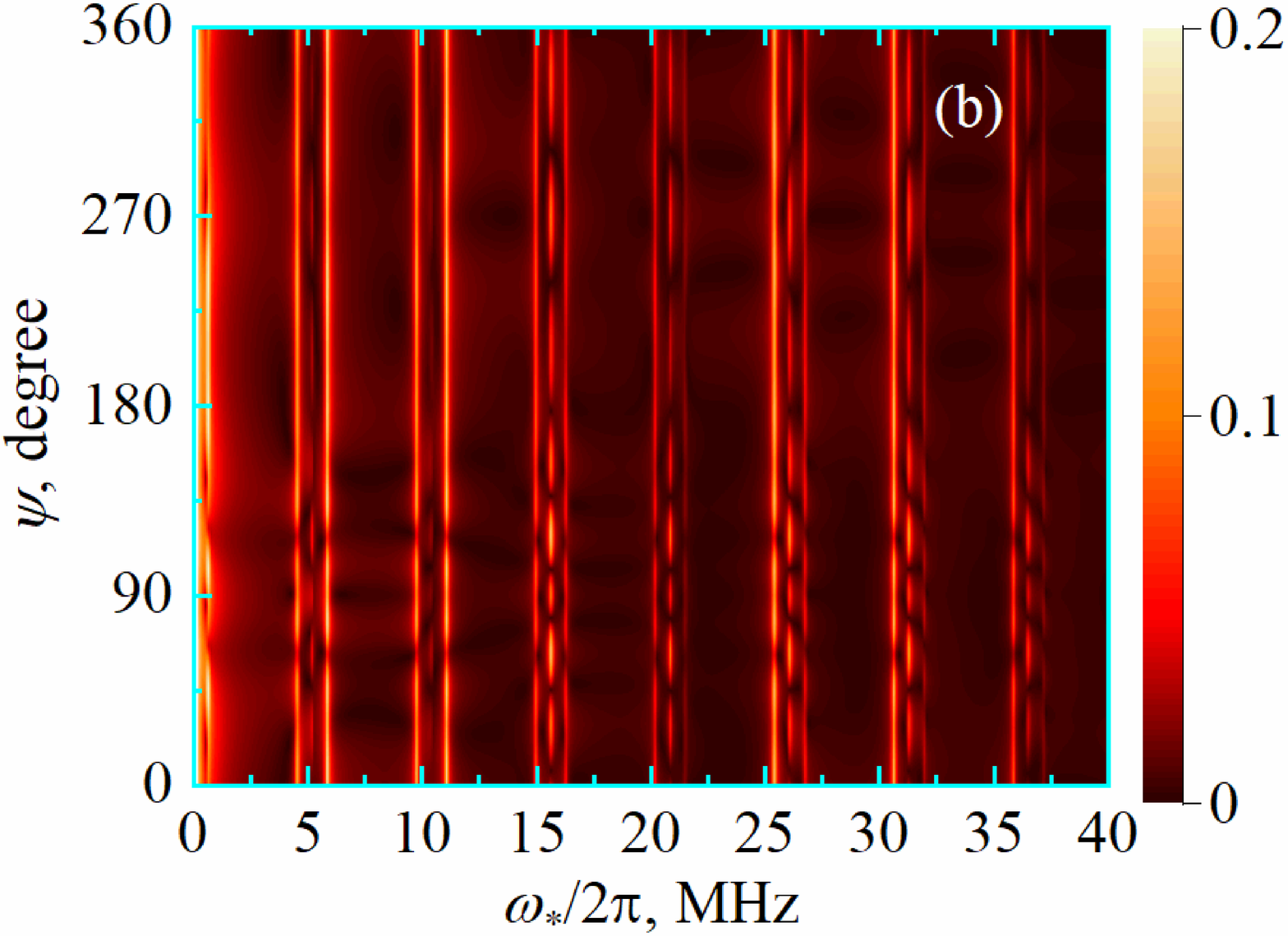}} \\
\end{minipage}
\caption{Fourier spectra as a function of the phase $\psi$ of the low-frequency driving field at $\Delta A/\omega$ equaled to 0 (a) and --0.25 (b). The other parameters are the same as in Fig. 3.
}
%%\label{fig4}
\end{figure*}
 We consider only the second-order Raman transition  when the resonance condition $\omega _{0}/2 =\omega$ is fulfilled.This transition is well observed and has the largest Rabi frequency among others multiphoton Raman transitions \cite{p47}. The Hamiltonian $H_{2}$ contains an infinite sum of oscillating harmonics with the frequencies which are integer multiples of the frequency $\omega $. There are no oscillations for $n=-1$ and $n=-3$. Therefore, the terms of the sum with these \textit{n} give the largest contribution and correspond to the RWA. At the strong coupling condition $0.1<A/\omega <1$ the other oscillating terms are significant. Their contribution  can be taken into account using the Bogoliubov averaging method \cite{ppBog} for constructing time-independent effective Hamiltonian in the framework of the non-secular perturbation theory. The averaging procedure up to the second order in $A\cos \theta /\omega $ (see \cite{ppBog,p49}) gives the following effective Hamiltonian: $H_{2} \to H_{eff} =H_{2}^{(1)} +H_{2}^{(2)} $, where
$$H_{2}^{(1)} =<H_{2} (t)>,$$
\begin{equation} \label{eq_03}
H_{2}^{(2)} =\frac{i}{2} <[\int _{}^{t}d\tau (H_{2} (\tau )-<H_{2} (\tau )>),H_{2} (t) ]>.
\end{equation}

Here the symbol $\langle . . . \rangle $ denotes time averaging over rapid oscillations of the type $\exp (\pm im\omega t)$ given by $\langle O(t)\rangle =\frac{\omega }{2\pi } \int _{0}^{{2\pi \mathord{\left/ {\vphantom {2\pi \omega }} \right. \kern-\nulldelimiterspace} \omega } }O(t)dt $. The upper limit \textit{t} of the indefinite integral indicates the variable on which the result of the integration depends, and square brackets denote the commutation operation.  As a result,  the  effective Hamiltonian can be written as

$$H_{eff} = \frac{\omega _{2}^{BS} }{2} \sigma ^{z} +\frac{\Omega _{2} }{2} (\sigma ^{+} e^{-i2\psi } +H.c.),$$ where
\begin{gather} \label{eq1}
\nonumber \Omega _{2} = 4\frac{J_{2} (a)}{a} A\cos \theta, \\
\nonumber \omega _{2}^{BS} = \frac{A^{2} \cos ^{2} \theta }{2\omega }\times \\
\times\left\{\sum _{n\ne -3}\frac{J_{n}^{2} +J_{n} J_{n+2} }{n+3} +\sum _{n\ne -1}\frac{J_{n}^{2} +J_{n} J_{n-2} }{n+1}\right\}.
\end{gather}

Here $\Omega _{2}$ is the Rabi frequency of the second-order Raman transition in the RWA and $\omega _{2}^{BS}$ is the Bloch--Siegert frequency shift for the second-order transition caused by the non-resonant rapidly oscillating non-RWA terms.  The Bessel function $J_{2} (a)$ in Eq.(4) for $\Omega _{2}$ appears due to virtual multiphoton transitions, in which the number of absorbed (emitted) photons exceeds by 2 the number of emitted (absorbed) photons. In the equation for $\omega_{2}^{BS}$ we omit the argument $a$ of the Bessel functions.

In the following, we will consider the experimental situation with a NV center when the microwave field excites transitions between the spin sublevels ${\left| 0 \right\rangle}$ and ${\left| -1 \right\rangle}$ of this center, while the level ${\left| +1 \right\rangle}$ is far detuned \cite{p47}. We assume that the spin system is initially in the ground state ${\left| 0 \right\rangle}$. Then, for the second-order Raman transition the probability to find the system in some moment again in the ground state $P_{{\left| 0 \right\rangle} }^{(2)} (t)$ is:
\begin{multline} \label{eq2}
P_{{\left| 0 \right\rangle} }^{(2)} (t) = \frac{1}{2} (1+\cos ^{2} \theta -2c_{1})+\\
+e\cos \left[2\omega t-a\cos (\omega t+\psi)-\phi_{e} \right]+\\
+c\cos (\Omega _{2}^{*} t-\phi _{c})+\\
+b\cos \Omega _{2}^{*} t\cos \left[2\omega t-a\cos (\omega t+\psi)-\phi _{b} \right]+\\
+d\sin \Omega _{2}^{*} t\cos \left[2\omega t-a\cos (\omega t+\psi)-\phi _{d} \right],
\end{multline}
where the effective Rabi frequency $\Omega _{2}^{*} = \sqrt{\Omega _{2}^{2} +(\omega _{2}^{BS})^{2} }$ is introduced taking into account the Bloch--Siegert shift and the following denotations are used: $c = (c_{1}^{2} +c_{2}^{2})^{1/2}$, $e = (e_{0}^{2} +e_{\pi /2}^{2})^{1/2}$, $b = (b_{0}^{2} +b_{\pi /2}^{2})^{1/2}$, $d = (d_{0}^{2} +d_{\pi /2}^{2})^{1/2}$;
 $\cos \phi _{c} = c_{1}/c$, $\cos \phi _{e} = e_{0}/e$, $\cos \phi _{b} = b_{0}/b$, $\cos \phi _{d} = d_{0}/d$;
\begin{gather}
 \nonumber c_{1} = \frac{\omega _{2}^{BS} \Omega _{2} }{4\Omega _{2}^{*2} } \sin 2\theta \cos \left(2\psi -a\cos \psi \right)+\frac{1}{2} \left(\frac{\Omega _{2} }{\Omega _{2}^{*} } \right)^{2} \cos ^{2} \theta, \\
\nonumber c_{2} = \frac{\Omega _{2} }{4\Omega _{2}^{*} } \sin 2\theta \sin (2\psi -a\cos \psi),
\end{gather}
\begin{multline}
\nonumber
e_{0} = \frac{1}{4} \left[\frac{\Omega _{2}^{2} }{\Omega _{2}^{*2} } \sin ^{2} \theta \cos (4\psi -a\cos \psi)\right.+\\
+\left(\frac{\Omega {}_{2} }{\Omega _{2}^{*} } \right)^{2} \sin ^{2} \theta \cos (a\cos \psi)-\\
-\left.\frac{\Omega _{2} \omega _{2}^{BS} }{\Omega _{2}^{*2} } \sin 2\theta \cos 2\psi \right],
\end{multline}
\begin{gather}
\nonumber b_{0} = -e_{0} +\frac{1}{2} \sin ^{2} \theta \cos (a\cos \psi), \\
\nonumber d_{0} = -\frac{\Omega _{2} }{4\Omega _{2}^{*} } \sin 2\theta \sin 2\psi -\frac{\omega _{2}^{BS} }{2\Omega _{2}^{*} } \sin ^{2} \theta \sin (a\cos \psi).
\end{gather}
We use labels $0$ and $\pi/2$ for $e$, $b$ and $d$. The expressions for $e_{\pi /2}, b_{\pi /2}, d_{\pi /2}$ are obtained from $e_{0}, b_{0}, d_{0}$ by the addition of the phase $\pi /2$ in the arguments  of sine and cosine functions  containing  $\psi$, for example, $ \cos (4\psi -a\cos \psi)\to \cos (4\psi -a\cos \psi +\pi /2)$, $ \cos 2\psi\to \cos (2\psi +\pi /2)$, and so on.

\section{TIME AND SPECTRAL MANIFESTATIONS OF THE BLOCH--SIEGERT EFFECT}

To demonstrate the possibility of direct observation of the Bloch--Siegert effect, we  calculate from Eq. \eqref{eq2} time and spectral features of the second-order Raman transition at the parameters of the driving field which can be realized in experimental studies similar to those in the experiment \cite{p47}.

Figure 1 shows the dependences of the RWA Rabi frequency $\Omega _{2}$, the non-RWA Rabi frequency $\Omega _{2}^{*}$ and the Bloch--Siegert-like shift $\omega _{2}^{BS}$ on the normalized driving strength $a = 2A\sin \theta /\omega$. The dashed line presents the non-RWA Rabi frequency $\tilde{\Omega }_{2}^{*}$ \cite{p49} taking into account the contribution of the third order in the parameter $A\cos \theta /\omega$; this contribution is very small.

When approaching a value of $A^{*} /\omega$, where $A^{*}/\omega = a^{*} /2\sin \theta$ and $a^{*}$ is the first root of the equation $J_{2} (a) = 0$ in the expression (4) for $\Omega _{2}$, the RWA Rabi frequency $\Omega _{2}$ tends to zero due to destructive interference of multiple photon processes, and the non-RWA Rabi frequency $\Omega _{2}^{*}$ reaches its minimum: $\Omega _{2}^{*} = \sqrt{\Omega _{2}^{2} +(\omega _{2}^{BS})^{2} } = \omega _{2}^{BS}$. This first minimum is at the first zero crossing of the Bessel function $J_{2} (a)$. The first zero is a $A/\omega = 2. 68$, the second zero is at $A/\omega = 4.39$, etc. The non-RWA Rabi frequency reaches its minima at all values of $A/\omega$ corresponding the zero crossing of the Bessel function.

The first term of Eq. \eqref{eq2} for $P_{{\left| 0 \right\rangle} }^{(2)} (t)$ is time-independent. This term as well as the values of the parameters $A, e, b$ and $d$ are a function of the physical parameters of the quantum system and the driving high- and low-frequency fields. Using decomposition by a series of the Bessel functions, one can see that the second term of Eq. \eqref{eq2} describes  rapid oscillations at the ``carrier'' frequencies $n\omega$, where $n$ is integer. The third term presents  slow oscillations at the non-RWA Rabi frequency $\Omega _{2}^{*}$. The fourth and fifth terms describe  oscillations at the ``carrier'' frequencies $n\omega$ with  amplitudes modulated by the Rabi oscillations at the frequency $\Omega _{2}^{*}$. Fig. 2 illustrates the evolution of $P_{{\left| 0 \right\rangle} }^{(2)} (t)$ for two values of the phase $\psi$ of the modulation field near and at the first minimum of the non-RWA Rabi frequency $\Omega _{2}^{*}$. The phase influences the oscillations of the ground state population of the drivev qubit and the strongest differences are observed between the oscillations at $\psi = 0$ and $\psi = 90^{0}$.

At the value of $A^{*} /\omega$ the RWA Rabi frequency $\Omega _{2} = 0$, the non-RWA Rabi frequency $\Omega _{2}^{*} = \omega _{2}^{BS}$, and $P_{{\left| 0 \right\rangle} }^{(2)} (t)$ can be written as
\begin{gather} \label{eq3}
P_{{\left| 0 \right\rangle} }^{(2)} (t;\Omega _{2} = 0) = \frac{1}{2} (1+\cos ^{2} \theta)+\\
\nonumber +\frac{1}{2} \sin ^{2} \theta \cos \left[\omega _{2}^{BS} t+2\omega t-a^{*} \cos (\omega t+\psi)+a^{*} \cos \psi \right].
\end{gather}

We see in Fig. 2 that at $\Delta A/\omega $ = 0 the amplitude modulation vanishes and the oscillations with the constant amplitude and periodically changing frequency occur. Thus, the disappearance of the amplitude modulation in the evolution of the ground state population is the evidence that the RWA Rabi frequency vanishes and the non-RWA Rabi frequency becomes equal to the Bloch--Siegert shift. In this case, the Bloch--Siegert oscillation is observed as the frequency modulation of the coherent response and is presented in Fig. 2 by the red line. The used parameters correspond to the ultrastrong regime with the coupling constant defined by $g/\varepsilon=Acos\theta/\omega\approx0.8$.

The disappearance of the RWA Rabi frequency represents a kind of electromagnetically induced transparency. The two-level system may become transparent under its bichromatic driving  when conditions of the RWA are fulfilled for the high- and low-frequency field \cite{p50, p51}. This effect is based on the destructive interference of multiple photon processes excited by the bichromatic driving; as a result, for properly chosen experimental parameters of the low-frequency field, amplitude of overall multiple photon transitions between the dressed states becomes zero. In contrast, at the strong driving when the non-RWA must be used for the low-frequency field, the full electromagnetically induced transparency cannot be realized, because the Bloch--Siegert oscillation remains if even the RWA Rabi frequency vanishes.

The Fourier spectra of $P_{{\left| 0 \right\rangle} }^{(2)} (t)$, $F(\omega _{*}) = \int _{0}^{\infty }e^{-i\omega _{*} t} e^{-\gamma t} P_{{\left| 0 \right\rangle} }^{(2)} (t) dt$, are shown in Fig. 3 for two values of the phase of the modulation field. The decay rate $\gamma$ was introduced phenomenologically using its value corresponding to a coherence time of 4 $\mu$s \cite{p47}. The spectra consist of Lorentzian lines at zero frequency, at the non-RWA Rabi frequency $\Omega _{2}^{*}$ and series of triplets are observed at frequencies $n\omega$ and $n\omega \pm \Omega _{2}^{*}$ corresponding to the amplitude-modulated oscillations. When $a\to a^{*}$ (or $A\to A^{*}$), the RWA Rabi frequency $\Omega _{2} \to 0$ as well as the coefficients $A_{1}, A_{2}, e_{0}, e_{\pi /2} \to 0$ and only the coefficients $d_{0}, d_{\pi /2}, b_{0}, b_{\pi /2}$ have non-zero values. In this case the line, corresponding to the oscillations at the RWA Rabi frequency, vanishes. At $A = A^{*}$, the non-RWA Rabi frequency $\Omega _{2}^{*} = \omega _{2}^{BS}$, only two side lines at the frequencies $n\omega \pm \omega _{2}^{BS}$ are remained in triplets and each triplet are transformed into doublet. A splitting between the doublet lines becomes exactly equal to $2\omega _{2}^{BS}$. Note that the fourth triplet near $\omega /2\pi$ = 20 MHz degenerates in a singlet. Indeed, using decomposition expansion of Eq. \eqref{eq3} by a series of the Bessel functions, the time-dependent part of this equation can be written as $\frac{1}{2} \sum _{n = -\infty }^{n = \infty }J_{n} (a^{*})\exp \left\{i\left[\left((n-2)\omega -\omega _{2}^{BS} \right)t\right.\right.$ $\left.\left.-a^{*} \cos \psi +n(\psi+\pi/2) \right]\right\} +c.c.$ It follows directly from this expression that the line with the higher frequency in the doublet ($4\omega -\omega _{2}^{BS}, 4\omega +\omega _{2}^{BS}$) at $n = -2$ and $n = 6$ vanishes, because its amplitude $J_{-2} (a^{*}) = 0$. It is an additional indication that the RWA Rabi frequency $ \Omega _{2}$ becomes equal to zero and the conditions are realized when the double Bloch--Siegert shift $2\omega _{2}^{BS}$ can  directly be determined from the splitting between the doublet lines in the spectrum $F(\omega_{*})$.

Features of the coherent oscillations in the qubit's evolution depend on the phase $\psi$ of the modulation field that results in the phase dependences of the spectral line amplitudes. Fig. 4 illustrates such dependences for $\Delta A/\omega$ equaled to 0 and --0.25. At $\Delta A/\omega = 0$, doublets occurs at the frequencies $n\omega \pm \omega _{2}^{BS}$ for all values of the phase of the modulation field; the spectra are practically phase-independent (Fig. 4(a)). On the other hand, at $\Delta A/\omega \ne 0$ the spectra depend on the phase of the low-frequency field and the strongest differences for the intensity of triplets are  at $\psi = 0$ and $\psi = 90^{0}$ (Fig. 4(b)).

\section{CONCLUSION}
We have studied the second-order Raman transition excited by the amplitude-modulated microwave field in the two-level system. It was shown that at the ultrastrong driving by the low-frequency modulation field the excited effective Rabi frequency is significantly modified by the Bloch--Siegert shift $\omega _{2}^{BS}$ due to multiphoton antiresonant interactions between the quantum system and the modulation field. In this regime, when the coupling strength $A$ exceeds the modulation frequency $\omega$, the RWA Rabi frequency $\Omega _{2}$ can become zero and the non-RWA Rabi frequency $\Omega _{2}^{*}$ is equal to the Bloch--Siegert shift $\omega _{2}^{BS}$. That results in the qualitative changes in the coherent dynamics of the system, i.e. the amplitude modulation of the oscillations in the evolution of the ground state population transforms into the oscillations of the constant amplitude and periodically changing frequency. In the Fourier spectra of the coherent response triplets are transformed into doublets with the splitting between the doublet lines equal to $2\omega _{2}^{BS}$. So, it is a unique possibility for the direct measuring the Bloch--Siegert shift in experiments on observation of  Raman transitions in  driven quantum systems. We demonstrate this possibility using calculations of the time and spectral behavior of the coherent oscillations in the qubit's evolution in the conditions of  experiments with the NV center in diamond, where  Raman transitions were observed \cite{p47}. Such possibility can also be realized for the high-order Raman transitions. Unlike the second-order transition, for the third- and fourth-order transitions the Bloch--Siegert shift can give the dominant contribution to the non-RWA Rabi frequencies [49]. However, the non-RWA Rabi frequency strongly decreases with increasing the transition order making difficulties for experimental studies of the high-order Raman transitions. The proposed direct observation of the Bloch--Siegert oscillations reveals additional features of coherent Raman dynamics and may be used as a new technique for studying quantum systems at bichromatic and multichromatic driving in the ultastrong regime.

\section{Acknowledgements}
The work was supported by State Programm of Scientific Investigations ``Physical material science, new materials and technologies'', 2016-2020, and by Project 01-3-1137-2019/2023, JINR, Dubna.

\end{document}